\documentclass[11pt,twoside]{article}

\usepackage{ihepconf}

\def\be{\begin{equation}}
\def\ee{\end{equation}}

\def\v{\varphi}

\input{amssym.def}
\input{amssym}

\input{amssym.def}
\input{amssym}
\begin{document}

  \title{CAN BLACK HOLES BE PRODUCED \\ [1mm]at HIGH-ENERGY COLLIDERS?}
 
 \author{{\bf V.A. Petrov}\\
 {\it Institute for High Energy Physics, 142200, Protvino,
 Russia}}

  \maketitle

$\bullet $ The present boom of interest to the black holes is very
much substantiated by accumulation of astrophysical observations,
which can be interpreted in terms of black holes~[1]. This,
certainly, refers to very large-mass systems that are to be
described in the framework of classical (non-quantum) field theory
(General Relativity or other classical theories of gravitation).

Quantum properties of the black holes (BH in the following) were
addressed some 30~years ago~[2] with a conclusion that quantum
effects lead to the thermal  radiation from the BH with an effective
``temperature'' inversely proportional to the BH mass. Quick
evaporation of the BH with small masses caused some hopes that the
``mini-BH'' may be relevant in the  high-energy collisions of
subnuclear particles, and the corresponding events could be
observable due to the characteristic event shapes. However expected 
sizes of such effects are negligible at present or near (LHC)
energies because of the small value of the Newton constant $\sim
M^{-2}_{PL}$. It is of interest to note that one of the most fervent
prophets of the black holes, S.W.~Hawking, has recently expressed
his deep doubts that the true event horizon ever forms at all 
(S.~Hawking's talk at GR 17, Dublin, 2004). 

$\bullet$ New prospects for exploration of the BH effects were opened
by the theories with extra-dimensions~[3] in which gravitational
interactions at small distances could get very strong at the scales
of order 1~TeV$^{-1}$.

The mechanism of the BH production in high-energy collisions is
usually presented as follows. 

Two energetic partons from the colliding hadrons (protons) with the
c.m.s. energy $\sqrt{s}$ --- considered as a component of the
energy-momentum tensor in the r.h.s. of the Hilbert--Einstein
equation~--- produce the metric perturbation corresponding to a BH of
the mass $\sqrt{s}$. This configuration having the Hawking
temperature $1/\sqrt s$ quickly evaporates giving rise to a specific
final multiparticle state. The cross-section of such elementary
process is assumed to be purely geometric
$$
\hat\sigma_{\mbox{\tiny BH}} (s)\sim r^2_{\mbox{\tiny BH}}(s),
$$
where
$$
r_{\mbox{\tiny BH}} (s) \sim 1 \mbox{TeV}^{-1} (\sqrt
s/\mbox{TeV})^{1/(1+\Delta)},
$$
is the BH radius and $4+\Delta$ is the dimensionality of space-time.
The rate of the BH events can be fantastic: 10~Hz!~[4]. 

At first sight the arguments are quite natural. However more thorough
investigation reveals quite serious theoretical problems. 

$\bullet$ As we all probably know, theoretical description and,
hence, model predictions for observed processes like those which
happen in high-energy particle collisions are based on the $S$-matrix
formalism. $S$-matrix relates idealized asymptotic states at
space-temporal infinity. Asymptotic states  are necessary for the
very interpretation of the field theory in terms of particles. It is
well known that the very notion of particle is problematic in generic
curved space-times~[5]. 

There are gravitational theories where the flat, pseudoeuclidean
space-time is a geometrical basis for any interaction, including the
gravitational one~[6], and which do not encounter with  difficulties
mentioned above. However, in these theories there are no black holes
at all.

$\bullet$ With a hope that such problems can be resolved in some way
in GR we try to see if there are some plausible conditions for the BH
formation in violent collisions. 

The process would happen when two high-energy particles enter
the interaction region and strong gravitational fields are generated
due to their high energy. Moreover one assumes that some ``trapping
surface'' would form which finally transforms to a black hole with
mass that is a finite fraction of the collision c.m.s. 
energy~[7]. We have to stress that such a ``trapping surface''
(analogous
to the Schwarzschild horizon surface) is not yet obtained as an
exact solution of the Hilbert-Einstein equations.

As BH are classical field configurations one can argue that in order
to be relevant to the realm of the high energy collisions some
conditions of ``quasiclassicity'' must be respected. In analogy with
QED one can ask for a large number of virtual quanta. One also has to
have space-time curvature small enough to evade the problems of
quantum-gravitational character. In ref.~[8] it was shown that these
necessary conditions are in general violated~[9]. 

$\bullet$ There is another remark concerning the BH formation. It is
known that in the comoving frame the collapse happens in a finite
time. On the other hand the distant observer will observe an
infinitely long process. From the point of view of the particle
detectors in, say, LHC this is namely such a situation. 

One could argue that big red shifts from the collapsing radiating
matter get big in an exponentially fast way and ``practically'' it
takes a
finite time to ``almost collapse''. At the moment the mechanism of
the would-be BH formation is too unclear to completely remove such an
objection.

$\bullet$ It is clear that at our present stage of knowledge the BH
is a classical field configuration. The BH quantum properties are
under an active investigation but none conclusive result is achieved.

Coming back to the S-matrix formalism one could have in mind the
following. 

Usually, in particle physics, one deals, when considering the
scattering and production processes, with Fock states representing
the asymptotic  states of particles with definite masses and
momenta. 

In principle one can choose any other representation of the CCR and
work with states where quantum field has a given functional value:
$$
\hat\varphi | \varphi (x)>=\varphi (x)|\varphi (x)>, 
$$
where $\varphi (x)$ is a classical field configuration. 

Then one could introduce the amplitude
$$
<\varphi | S|p_1,p_2> \ ,
$$
which corresponds to the ``production'' of a field configuration
$\varphi (x)$ in the course of the collision of particles $p_1$ and
$p_2$.

In order to detect such a configuration with detectors that are
adapted to ``usual'' asymptotic states $|k_1,...k_n>$ one can use the
amplitude
$$
<k_1..k_n|\varphi >
<\varphi |S|p_1,p_2>.
$$
An important ingredient is the amplitude 
$<k_1...k_n|\varphi>$.  In
simple cases it can be easily computed. For instance, in the case
of single neutral scalar field one has~[10]:
$$
<0|\varphi>\sim \exp \left [
-\int d\vec x \v (\vec x) 
\sqrt{m^2-\vec\nabla^2_x}\v (\vec x)\right ],
$$
$$
<k_1|\v>
\sim
\tilde\v (\vec k_1)<0|\v>,\;\;\; \mbox{etc.}
$$
In our case $\v(x)$ is the Schwarzschild solution
$g^{Schw}_{\mu\nu}$:
$$
g^{Schw}_{\mu\nu} (r)
\begin{array}{c}
\Bigl | \Bigr.\\
\mbox{Hilbert's gauge}
\end{array} =
-\left (
1-\frac{r_{\mbox{\tiny BH}}}{r}
\right )
dt^2+
\left (
1-\frac{r_{\mbox{\tiny BH}}}{r}
\right )^{-1}
dr^2+r^2d\Omega^2,
$$
$$
\mbox{at}\;\; \Delta\neq 0\;\; r_{BH}/r\to (r_{BH}/r)^{1+\Delta}.
$$
In other words one has to compute the amplitude
$$
<g^{Schw}|S|
p_1,p_2>,
$$
where the state $<g^{Schw}|$ can be mixed with other representations.
Thus, the amplitude for production of the final
state $|k_1..k_n>$ which is a result of the BH ``decay'' could have
the following form:
$$
T^{BH} (k_1...k_n|p_1p_2)
=\sum_{m}<k_1...k_2|g^{Schw}>
<g^{Schw} | q_1...q_m>
<q_1...q_m|T|p_1p_2>\times
$$
$$
\times
(2\pi)^4\delta \left (\sum^{m}_{1} q_j-p_1-p_2\right).
$$
Note that $T$ contains the (quantum) gravitational interactions. 

There is, however, a big problem: in such a construction the
energy-momentum conservation is not guaranteed, generally. The state
$|g^{Schw}>$ is not an eigenstate of the energy-momentum operator.

$\bullet$ Thus, I conclude that in spite of quite exciting and
seemingly
natural theoretical prospects for observation of copious BH
production at the LHC, one encounters, at more close inspection,  
with lot of theoretical problems, including a proper definition of
corresponding amplitudes.

In general, one has to realize that if LHC experiments will not see
mini-BHs it will not  shake the  fundamental basis of our present worldview,
just show up some not very well grounded model assumptions. 

It does not mean that the very subject is hopeless  and devoid of
interest. Quite contrary to that!

\end{document}